\documentclass[aps,prl,amssymb,reprint,twocolumn,superscriptaddress,longbibliography]{revtex4-1}
\usepackage[ansinew]{inputenc}
\usepackage{graphicx}
\usepackage[breaklinks=false,colorlinks=true,linkcolor=blue,urlcolor=blue,citecolor=blue]{hyperref}
\usepackage{setspace}
\usepackage[ansinew]{inputenc}
\usepackage{caption}
\usepackage[labelfont=bf]{caption}
\usepackage{graphicx}
\usepackage{amsmath}
\usepackage{color}
\usepackage{lineno}
\usepackage{ulem}
\usepackage{subfigure}
\begin{document}

%Title of paper
\title{Compact realization of all-attosecond pump-probe spectroscopy} 

\author{M. Kretschmar} 
\affiliation{Max-Born-Institut, Max-Born-Strasse 2A, 12489 Berlin, Germany}
\author{E. Svirplys}
\affiliation{Max-Born-Institut, Max-Born-Strasse 2A, 12489 Berlin, Germany}
\author{M. Volkov}
\affiliation{Max-Born-Institut, Max-Born-Strasse 2A, 12489 Berlin, Germany}
\author{T. Witting}
\affiliation{Max-Born-Institut, Max-Born-Strasse 2A, 12489 Berlin, Germany}
\author{T. Nagy}
\affiliation{Max-Born-Institut, Max-Born-Strasse 2A, 12489 Berlin, Germany}
\author{M. J. J. Vrakking}
\affiliation{Max-Born-Institut, Max-Born-Strasse 2A, 12489 Berlin, Germany}
\author{B. Sch\"utte}
\email{Bernd.Schuette@mbi-berlin.de}
\affiliation{Max-Born-Institut, Max-Born-Strasse 2A, 12489 Berlin, Germany}

\date{\today}

\begin{abstract}
The ability to perform attosecond-pump attosecond-probe spectroscopy (APAPS) is a longstanding goal in ultrafast science. While first pioneering experiments demonstrated the feasibility of APAPS, the low repetition rates ($10-120$~Hz) and the large footprints of existing setups have so far hindered the widespread exploitation of APAPS. Here we demonstrate two-color APAPS using a commercial laser system at 1~kHz, straightforward post-compression in a hollow-core fiber and a compact high-harmonic generation (HHG) setup. The latter enables the generation of intense extreme-ultraviolet (XUV) pulses by using an out-of-focus HHG geometry and by exploiting a transient blueshift of the driving laser in the HHG medium. Near-isolated attosecond pulses are generated, as demonstrated by one-color and two-color XUV-pump XUV-probe experiments. Our concept allows selective pumping and probing on extremely short timescales and permits investigations of fundamental processes that are not accessible by other pump-probe techniques.
\end{abstract}
\maketitle

The development of attosecond science~\cite{krausz09} at the beginning of this century has provided unprecedented opportunities for the study of electron dynamics in atoms~\cite{uiberacker07}, molecules~\cite{lepine14}, liquids~\cite{jordan20}, solids~\cite{schultze13} and at the nanoscale~\cite{ciappina17}. Up to now, the light source most often used in attosecond science is high-harmonic generation (HHG). Since HHG is typically limited to the generation of weak XUV pulses with very low on-target pulse energies (typically fJ - nJ), in the vast majority of experiments, it has not been possible to configure an attosecond-pump attosecond-probe experiment. Instead, attosecond pulses have been used in combination with intense near-infrared (NIR) pulses. Popular pump-probe schemes include attosecond streaking~\cite{itatani02, kienberger04} and attosecond transient absorption spectroscopy~\cite{geneaux19}, which have led to exciting insights into, among other things, photoionization time delays~\cite{schultze10}, valence electron motion~\cite{goulielmakis10} and the decay of inner-shell holes in atoms~\cite{drescher02}. 

%It is fair to say that the enormous progress in attosecond science in the past two decades has only been possible because of the numerous contributions which have been made in many laboratories around the world.

The use of intense NIR pulses as the pump or probe pulse in attosecond pump-probe experiments imposes several limitations, in particular for the following three reasons: (1) The time resolution that is achieved in a pump-probe experiment is normally given by the cross-correlation of the pump and probe pulses. For an XUV-NIR pump-probe sequence this cross-correlation is at least several femtoseconds, i.e. not in the attosecond domain. To circumvent this problem, attosecond science has adopted experimental protocols where the NIR optical period is used as a clock with attosecond time resolution~\cite{krausz09}. However, while these protocols are suitable for investigations of NIR-field-driven dynamics or electron dynamics that occurs on a sub-cycle timescale, these methods are typically not capable of resolving dynamics that extends from the attosecond to the few-femtosecond domain. (2) Since the on-target pulse energies of attosecond light sources generated by HHG are small, the signal levels in attosecond pump-probe experiments are small as well. As a result, the intensity of the NIR pulse that is used as the pump or the probe pulse in the pump-probe sequence often needs to be at a level where multi-photon interactions and field-driven modifications of the sample cannot be ruled out, potentially masking or changing the dynamics of interest. (3) NIR pulses particularly interact with valence-shell electrons which are not strongly localized around the nuclei in a sample. Therefore, NIR pulses are not very suitable for initiating or probing localized electron dynamics. By contrast, XUV and X-ray pulses strongly interact with core- and inner-valence shell electrons, thus providing opportunities for highly localized element-specific initiation and observation of electron dynamics. 

All these limiting factors associated with the use of NIR pulses within an attosecond pump-probe sequence can be overcome by using attosecond pulses in the XUV or X-ray range as the pump and the probe. Up to now, only a few attosecond-pump attosecond-probe spectroscopy (APAPS) experiments have been reported. In Ref.~\cite{tzallas11} Tzallas \textit{et al.} reported electronic coherences resulting from the preparation of a superposition of autoionizing states of Xe using an intense XUV pulse with an intensity envelope of 1.5~fs and making use of double Mach-Zehnder interferometric polarization gating. Subsequently, Takahashi $\textit{et al.}$ generated intense isolated attosecond pulses using a combination of driving pulses at 800~nm and 1300~nm, and measured a pulse duration of 500~as using an auto-correlation in nitrogen~\cite{takahashi13}. In addition, Midorikawa and co-workers have applied intense attosecond pulse trains to study electronic coherences in nitrogen~\cite{okino15} and in the dissociation of hydrogen molecules~\cite{nabekawa16}. In our laboratory we have recently applied intense few-femtosecond-long attosecond pulse trains generated using a driver based on optical parametric chirped-pulse amplification~\cite{kretschmar20} to study multi-electron dynamics in Ar~\cite{kretschmar22}. A very significant recent development has been the demonstration of attosecond pulse production at the Linac-Coherent Light Source (LCLS) in Stanford~\cite{duris20}. In a first time-resolved application of this source, coherent electronic motion in the course of Auger-Meitner decay was revealed, using a variant of the attosecond streaking technique~\cite{li22}. Recently, a first attosecond-pump attosecond-probe experiment in the soft X-ray range was reported as well~\cite{driver22}. 

All the previously reported APAPS experiments were performed using large setups, which are vulnerable to instabilities. Furthermore, highly specialized and low-repetition rate laser systems were used, the latter limiting the signal-to-noise ratio that could be obtained. For instance, whereas pump-laser-induced signal changes on the order of $10^{-4}$ could be recorded in several previous XUV-NIR pump-probe experiments~\cite{weisshaupt17, niedermayr22}, successful APAPS experiments to date relied on pump-laser-induced signal changes of $10 - 100$~$\%$~\cite{tzallas11, takahashi13, kretschmar22}. The combination of the above-mentioned factors has prevented the widespread application of APAPS. In order to enable the applicability of the technique in a large range of laboratories, a different concept is required, which improves the stability and statistics of the experiments.

\begin{figure}[tb]
 \centering
 \includegraphics[width=8cm] {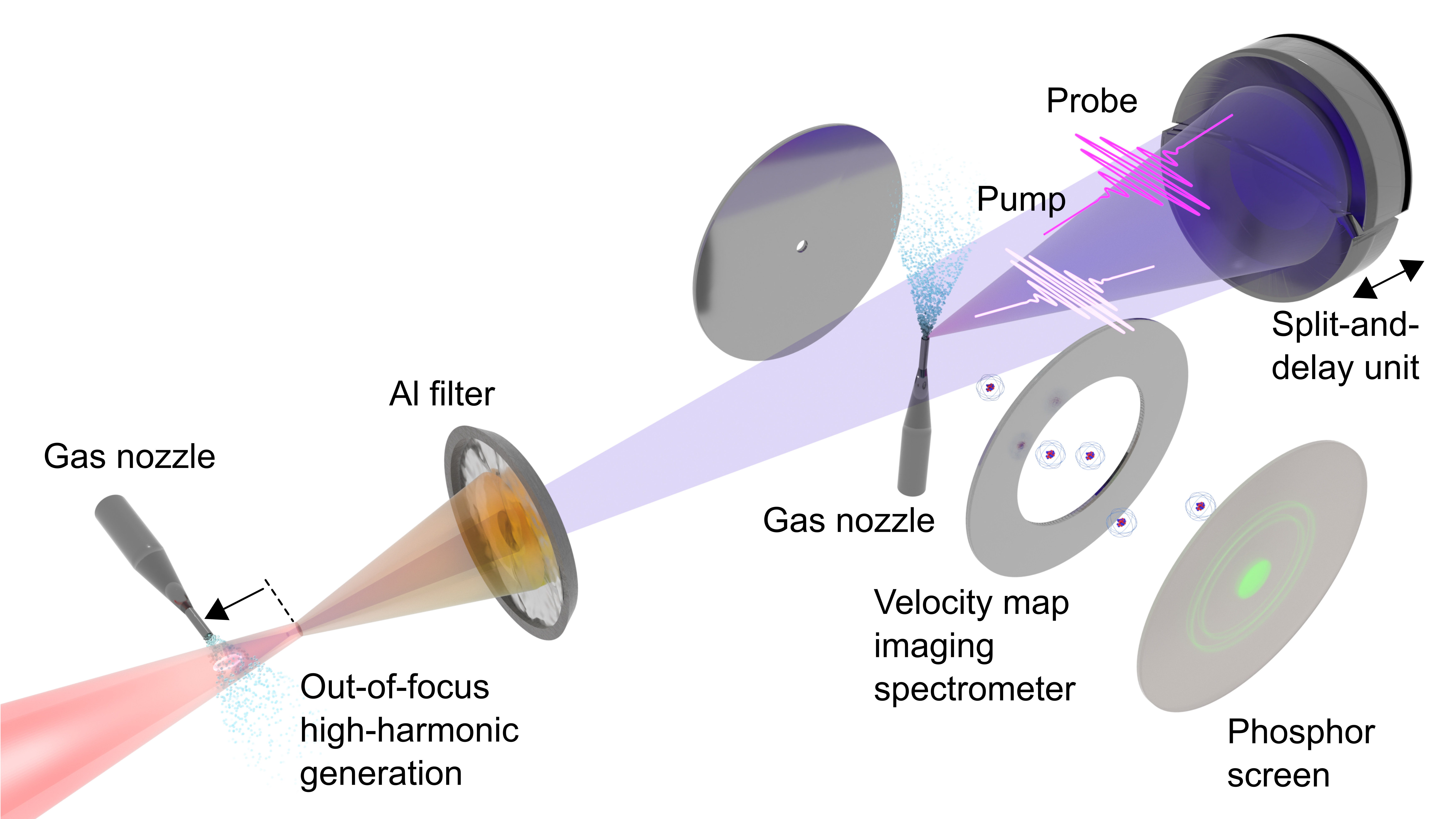}
 \caption{\label{figure_setup} \textbf{Experimental setup.} NIR pulses with a duration of 3.8~fs are focused into a vacuum chamber using a spherical mirror with a focal length of 75~cm. A pulsed gas jet (Kr or Xe at a backing pressure of 4~bar) is placed about 3.5~cm in front of the driving laser focus, and a 100-nm-thick Al filter is used to block the NIR beam. At a distance of 1.5~m behind the HHG jet, spherical multilayer mirrors with a focal length of 5~cm are used to spectrally select and focus the pump and the probe pulses. The generated ions are recorded using a velocity-map imaging spectrometer (VMIS).}
\end{figure}

\begin{figure}[htb]
 \centering
  \includegraphics[width=8.6cm] {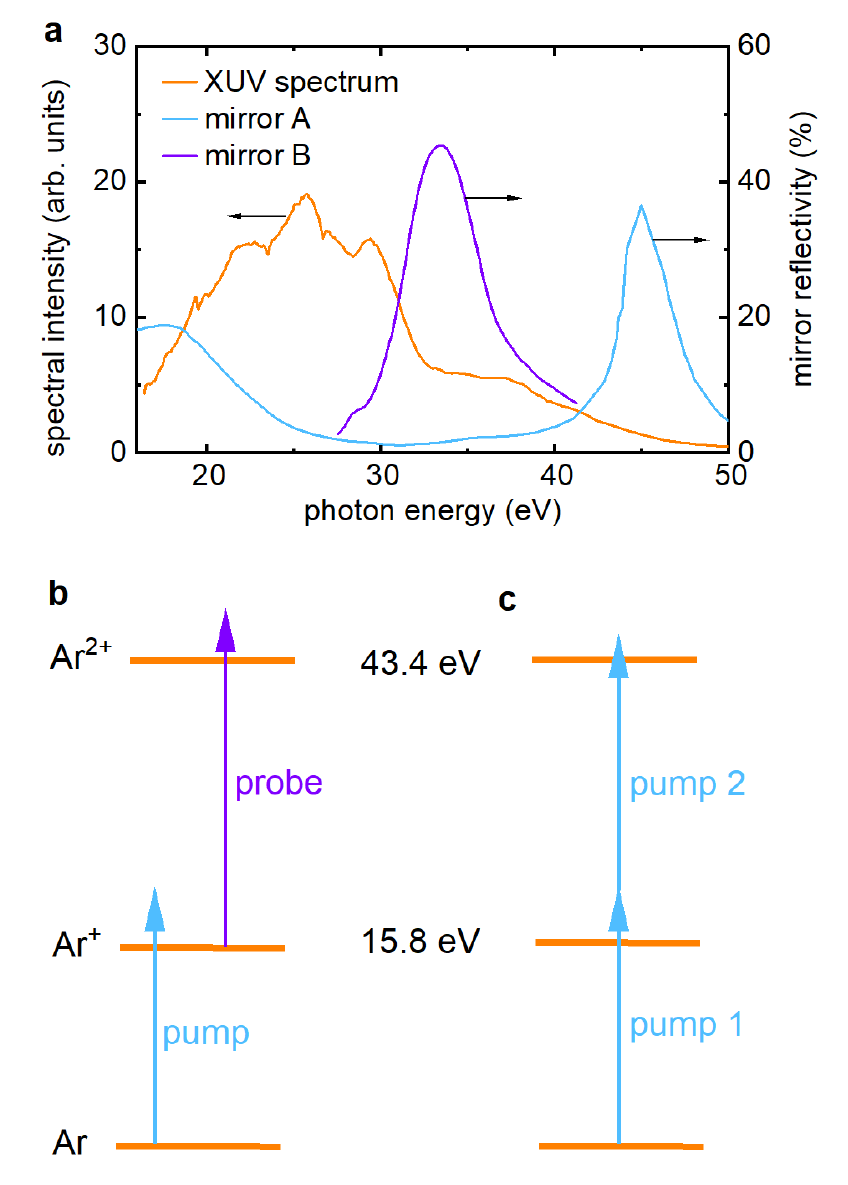}
 \caption{\label{figure_spectrum} \textbf{XUV spectrum and two-photon ionization scheme.} \textbf{a}, Measured XUV spectrum (orange curve) and simulated reflectivities of mirror A (blue curve) and mirror B (violet curve). The combination of the XUV spectrum and the XUV mirror reflectivity produces a pump spectrum that is centered around 20~eV (see Supplementary Information).\textbf{b}, Schematic of two-photon double ionization of Ar, where the first and the second ionization potential are indicated. Sequential ionization is possible using a pump pulse centered around 20~eV for the ionization of neutral Ar and a probe pulse at higher photon energies for the ionization of Ar$^+$. \textbf{c}, Two-photon two-electron ionization using two copies of the same pulse. }
\end{figure}

Here we pursue and successfully demonstrate the following strategy for the implementation of APAPS: (1) In order to be able to operate the experiment at a sufficiently high repetition rate and to be able to work with a proven and mature laser technology, we base the experiment on a commercial Ti:sapphire laser system operating at 1~kHz. Moreover, to keep the operation of the laser system as robust and simple as possible, we operate the system without carrier-envelope phase (CEP) stability. (2) The typical pulse duration of Ti:sapphire laser amplifiers is $\geq 15$~fs. Therefore, the generation of isolated attosecond pulses requires an efficient gating scheme during the HHG and / or post-compression of the laser prior to the HHG. Avoiding gating of the HHG, which is typically accompanied by reduced HHG conversion efficiencies, and avoiding the challenge of post-compressing high-energy laser pulses ($>5$~mJ)~\cite{nagy21}, a modest NIR pulse energy of 3~mJ is used. This enables straightforward spectral broadening in a hollow-core fiber and temporal compression down to 3.8~fs (see Fig.~\ref{figure_setup}) with a pulse energy of 1~mJ, see Methods for details. Similar specifications have been reached in a number of laboratories around the world, but have so far not led to the demonstration of APAPS. (3) To reach the high XUV intensities that are needed for APAPS, we make two significant changes in the way that the HHG process is implemented: (a) High harmonics are generated in a compact setup, where the driver laser is tightly focused to an intensity of $6 \times 10^{15}$~W/cm$^2$, and a high-density gas jet is placed several Rayleigh lengths in front of the driving laser focus. As a result, a relatively high XUV pulse energy of 10~nJ and a small virtual source size are obtained~\cite{major21}, enabling the generation of intense attosecond pulses upon refocusing of the diverging harmonics. (b) By performing HHG in an overdriven regime~\cite{johnson18}, the driving laser frequency is transiently up-shifted due to ionization-induced self-phase modulation (SPM) in the HHG medium~\cite{major23}. Since the HHG efficiency in Kr scales as about $\lambda_{driver}^{-6.5}$ with the driving laser wavelength $\lambda_{driver}$~\cite{shiner09}, the SPM-induced blueshift favorably influences the HHG efficiency and the achievable XUV intensity. (4) In previous APAPS schemes based on HHG, two copies of the attosecond pulses were generated, thereby preventing a clear distinction between the pump and the probe pulse. Here we demonstrate a two-color APAPS experiment, which is achieved by using two different spherical multilayer mirrors that select different photon energy ranges from the broadband attosecond pulse (see Fig.~\ref{figure_spectrum}\textbf{a}) and that tightly focus the XUV pulses to a small focus with a beam waist radius of about 1~$\mu$m. We note that one previous step towards two-color APAPS was the demonstration of attosecond pulses at two different photon energies from two consecutive gas jets~\cite{fabris15}. However, no APAPS results were reported from this setup.

\section{Results}

\subsection{Signal considerations in APAPS}

In order to assess the viability of the approach that we have described in the introduction, we will first estimate the number of two-photon ionization events events that can be achieved in a two-color pump-probe experiment in Ar where a pump pulse centered at 20~eV (cf. Fig.~S1\textbf{b}) ionizes neutral Ar producing an Ar$^+$ ion and where a time-delayed probe pulse centered at 33.5~eV further ionizes these Ar$^+$ ions and generates Ar$^{2+}$ ions (see Fig.~\ref{figure_spectrum}\textbf{b} for a scheme of the ionization process). In doing so, both the pump and the probe laser are assumed to follow Gaussian optics, and in agreement with the experiment it is assumed that the width of the atomic beam ($d$) is large compared to the Rayleigh length $z_R$ of the two XUV beams. Under these assumptions we derive (see Supplementary Information) that the number of Ar$^{2+}$ ions that are produced per laser shot is equal to 
\begin{equation}
    N_{event}= c \times \frac{2\pi n_0 N_{pump} \sigma_{pump} N_{probe} \sigma_{probe}}{ \lambda_{pump}},
\end{equation}
where $c=1$ if $\lambda_{pump}\gg\lambda_{probe}$ and $c=0.5$ if $\lambda_{pump}=\lambda_{probe}$. In our experiment neither of these conditions is fulfilled, therefore $0.5<c<1$. $n_0$ is the atomic density in the interaction region, $N_{pump}$ and $N_{probe}$ are the number of XUV photons in the pump and the probe beams, and $\sigma_{pump}$ and $\sigma_{probe}$ are the cross sections for the ionization of Ar and Ar$^+$ by the pump and the probe pulses, respectively. $\lambda_{pump}$ denotes the pump wavelength. Based on estimations and measurements in our experiment, we assume $n=10^{13}$~cm$^{-3}$, $N_{pump}=3.75 \times 10^{7}$ and $N_{probe}=2.8 \times 10^{7}$ photons. In addition, a cross section of 32.5~Mb is used for the ionization of Ar at 20~eV~\cite{chan92}, and a cross section of 10~Mb for the ionization of Ar$^+$ at 33.5~eV~\cite{covington11}. With these numbers, we predict the generation of 2-3 Ar$^{2+}$ ions per laser shot. While the above derivation was made for two collinear beams, two non-collinear beams were used in the experiment, for which we estimate that the number of Ar$^{2+}$ ions is about a factor of 3 lower. This estimation shows that under these conditions an APAPS experiment is feasible, as will be demonstrated in the next section.

\subsection{Two-color attosecond-pump attosecond-probe spectroscopy}

\begin{figure}[htb]
 \centering
  \includegraphics[width=8.6cm] {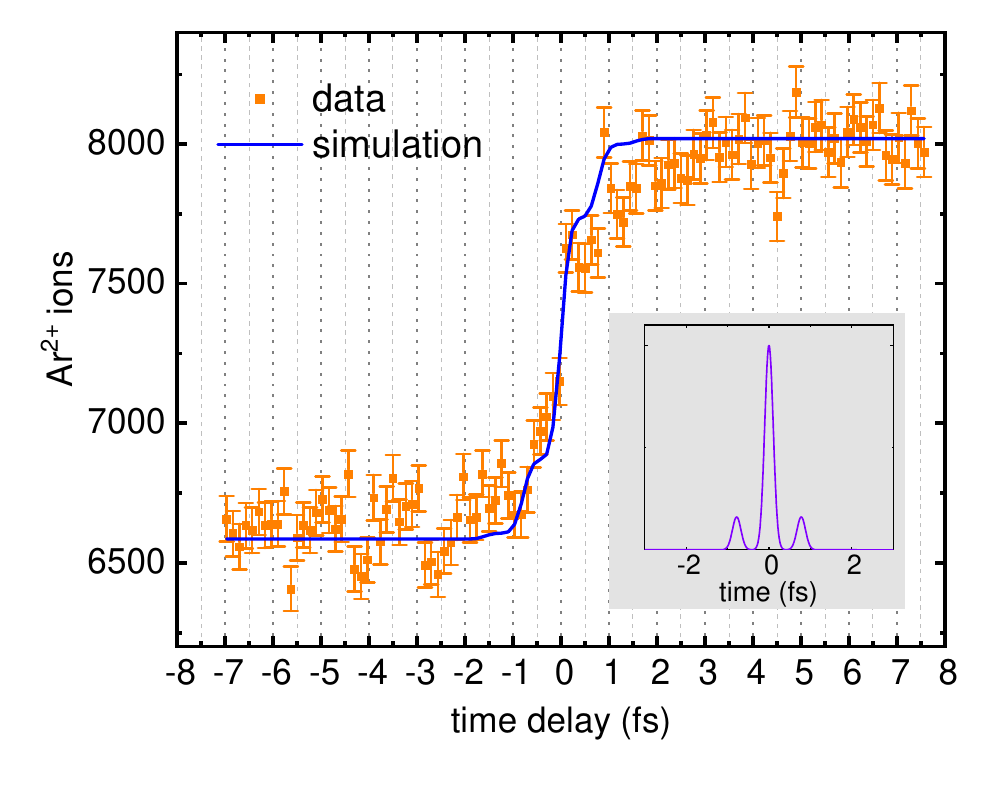}
 \caption{\label{figure_CC} \textbf{Two-color APAPS.} The generation of Ar$^+$, as induced by a broadband pump pulse, is probed by a second pulse with photon energies above the second ionization potential of Ar, thus producing Ar$^{2+}$. The error bars correspond to the shot noise error calculated from the estimated number of ions. The increase of the Ar$^{2+}$ ion yield around zero delay is explained by the more efficient generation of Ar$^{2+}$ when the probe pulse follows the pump pulse. The blue curve shows the results of a simultaneous fit of the cross-correlation and the auto-correlation (see Fig.~\ref{figure_AC}), giving pulses as shown in the inset. These pulses have a width of 240~as as well as pre- and post pulses with relative intensities of 16~$\%$.}
\end{figure}

In a first experiment, sequential two-photon Ar double ionization was studied as a function of the delay between an attosecond pump pulse at 20~eV and an attosecond probe pulse at 33.5~eV. To this end, two different multilayer mirrors (referred to as mirror A and mirror B) were used, selecting different spectral ranges for the pump and the probe pulses (see Fig.~\ref{figure_spectrum}\textbf{a}). Mirror A used for the pump pulse had a high reflectivity at photon energies below $25$~eV (and another reflectivity peak at 45~eV which played a negligible role in the current experiments, see Supplementary Information), whereas mirror B used for the probe pulse had a reflectivity peak at 33.5~eV. In the experiment, the pump pulse efficiently ionized Ar (which has an ionization potential (IP) of 15.8~eV). The formation of Ar$^+$ ions was then probed by further ionization leading to Ar$^{2+}$ using the probe pulse, requiring a photon energy of at least 27.6~eV. 

As shown in Fig.~\ref{figure_CC}, the Ar$^{2+}$ ion yield (orange data points) increases when the probe pulse arrives after the pump pulse (positive time delays). These data were recorded by scanning the pump-probe time delay twice, on each occasion accumulating Ar$^{2+}$ ions over 4000 laser shots. In average, the generation of 1 Ar$^{2+}$ ion / laser shot was estimated, in reasonable agreement with the afore-mentioned prediction. Measuring approximately 8000 Ar$^{2+}$ ions per delay point introduces a shot noise error of about 1~$\%$ in the data, in good agreement with the data that are shown in Fig.~\ref{figure_CC}. To exploit the full potential of APAPS, strategies to substantially reduce shot noise are discussed in the Discussion section.

The increase of the Ar$^{2+}$ ion yield takes place on an extremely fast timescale. As shown in Fig.~\ref{figure_CC} (blue curve), the Ar$^{2+}$ yield as a function of delay can be well reproduced by a model that considers the Ar$^{2+}$ as arising from ionization by a sequence of attosecond pulses. The pulses obtained from a fit of the experimental data (taking into account both the data from the cross-correlation measurement (Fig.~\ref{figure_CC}) and from an auto-correlation measurement (Fig.~\ref{figure_AC}) as discussed in the next section) are shown in the inset of Fig.~\ref{figure_CC}. They have pre- and post-pulses with a relative intensity of $(16\pm2)$~$\%$ each, and the width of an individual attosecond burst is 240~as. Accordingly the risetime of the step in the Ar$^{2+}$ yield as a function pump-probe delay, defined as the difference between the times where the Ar$^{2+}$ yield has increased by 20~$\%$, respectively 80~$\%$, is only 630~as.

\subsection{XUV auto-correlation}

\begin{figure}[tb]
 \centering
  \includegraphics[width=8.6cm] {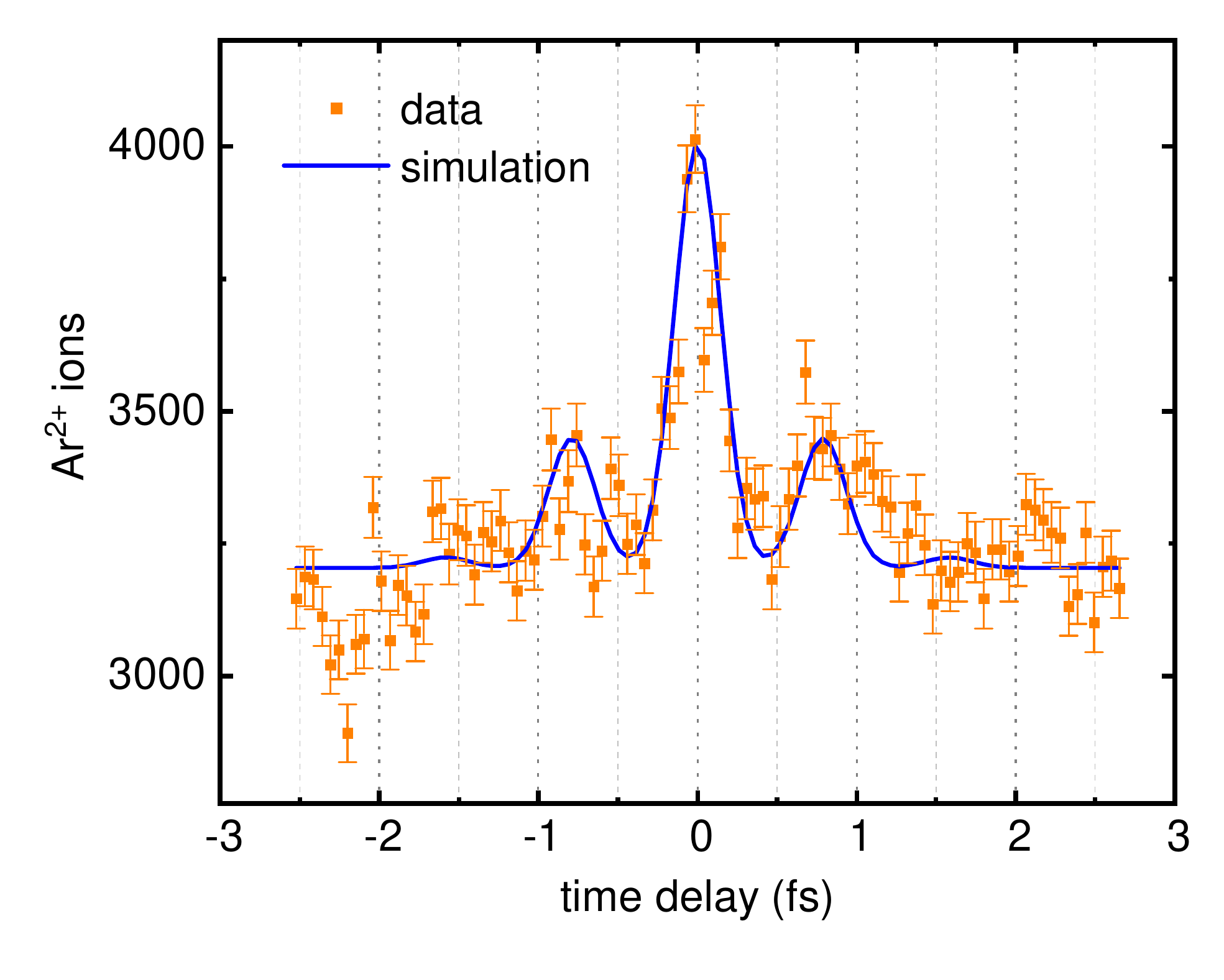}
 \caption{\label{figure_AC} \textbf{XUV auto-correlation.} Auto-correlation measurement of two identical XUV pulses in Ar, resulting in two-photon double ionization of Ar. The blue curve shows a simulation using the same attosecond pulses that were used in the simulation in Fig.~\ref{figure_CC} and that are shown in the inset of this figure.}
\end{figure}

In a second experiment, an auto-correlation measurement was performed based on the simultaneous absorption of two XUV photons, leading to the direct emission of two electrons and the formation of Ar$^{2+}$ (see Fig.~\ref{figure_spectrum}\textbf{c}). Therefore, this experiment is similar to the measurements performed in Refs.~\cite{tzallas11, takahashi13, kretschmar22}. Since the efficiency of simultaneous two-photon absorption depends on the XUV peak intensity, it can be controlled by varying the time delay between two attosecond pulse replicas, and the highest Ar$^{2+}$ ion yield is expected at zero time delay. To this end, two identical mirrors (of type A, meaning that the interaction is dominated by photons with energies $<25$~eV) were used to create an XUV pulse pair with a variable delay. 

The result is presented in Fig.~\ref{figure_AC} (orange data points), showing a dominant peak at zero delay. As before, these data were recorded by scanning the pump-probe time delay twice, on each occasion accumulating Ar$^{2+}$ ions over 4000 laser shots. In this measurement, the estimated number of Ar$^{2+}$ ions per shot was about 0.5, giving a shot noise error of about $2$~$\%$. The blue curve shows a fit using the same attosecond pulse that was used for the simulation shown in Fig.~\ref{figure_CC}. Small additional peaks are visible around $-0.8$~fs and $0.8$~fs delay, which are attributed to the attosecond pre- and post-pulses as indicated in the inset of Fig.~\ref{figure_CC}. Given that the presence / absence of multiple attosecond pulses is known to depend strongly on the CEP, the modest relative intensity of the attosecond pre- and post-pulses suggest that for certain CEP values an isolated attosecond pulse may be generated. In the future, we envision performing single-shot measurements in combination with single-shot CEP tagging, allowing to sort the data with respect to different CEP intervals. This will require an increase of the count rate in the experiment. Strategies how to achieve this will be presented in the Discussion section.

\begin{figure}[htb]
 \centering
  \includegraphics[width=8.6cm] {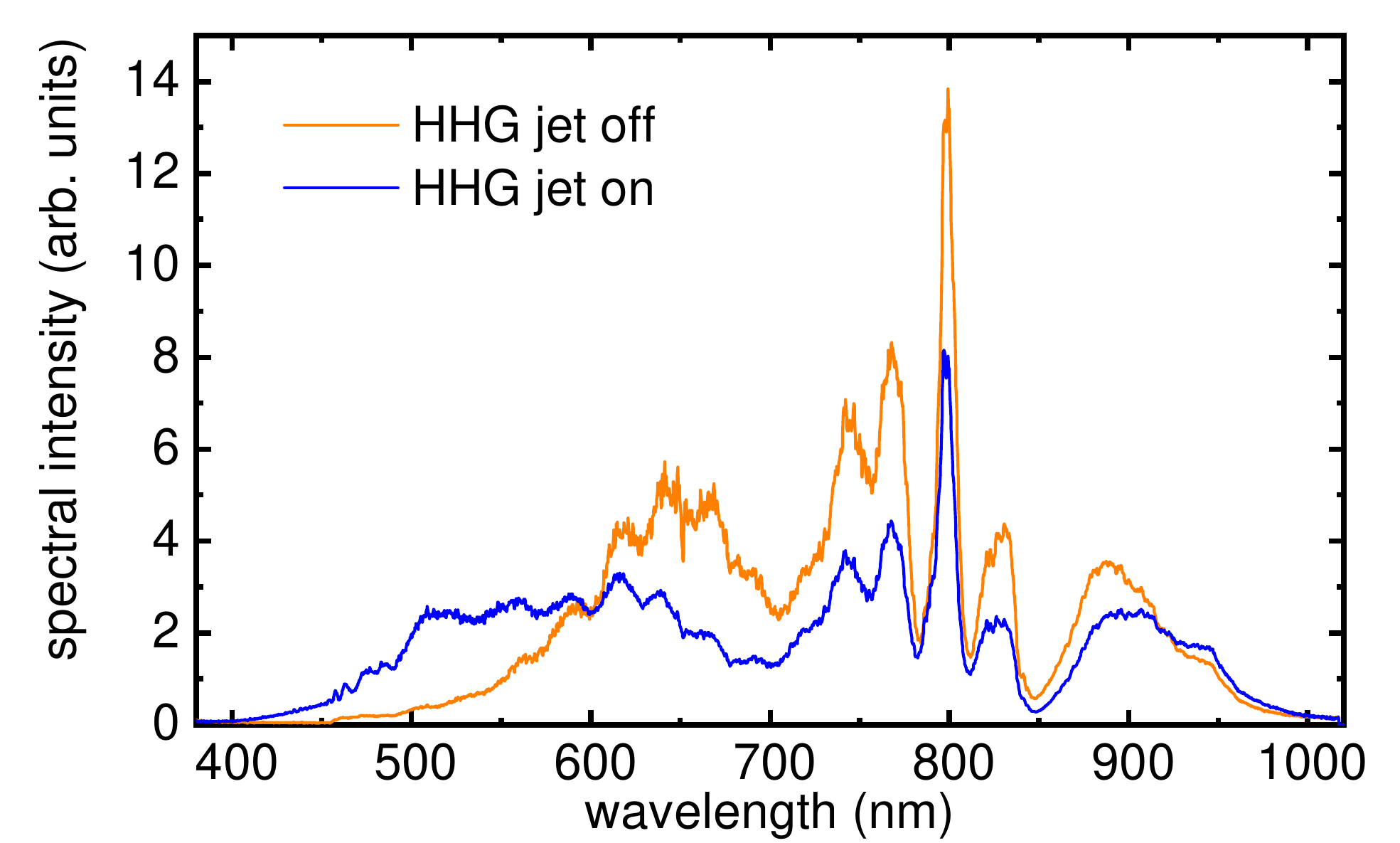}
 \caption{\label{figure_blueshift} Driving laser spectrum behind the gas jet used for HHG in Kr when the jet is switched off (orange curve) and when the jet is switched on (blue curve), showing a substantial blueshift in the HHG medium, which is attributed to ionization-induced SPM.}
\end{figure}

Surprisingly, the temporal separation of the peaks in Fig.~\ref{figure_AC} is significantly shorter than half the oscillation period of the driving laser ($\approx 1.3$~fs). Therefore, the experimental results suggest that the HHG was driven by a laser field characterized by a central wavelength substantially shorter than the central wavelength of the post-compressed Ti:sapphire laser. We attribute this to a blueshift of the driving laser in the HHG medium prior to the onset of efficient HHG. This interpretation is further supported by the experimentally measured driving laser spectra shown in (Fig.~\ref{figure_blueshift}). The spectrum after passage through the HHG jet (blue curve) is substantially blue-shifted with respect to the spectrum that is measured when the HHG jet is switched off (orange curve). The blueshift is attributed to ionization-induced self-phase modulation~\cite{bloembergen73, major23} and influences the HHG conversion efficiency, which was previously experimentally shown to scale as $\lambda_{driver}^{-6.5}$ for HHG in Kr~\cite{shiner09}, where $\lambda_{driver}$ represents the driving laser wavelength.Thus the ionization-induced blueshift is an important contribution to being able to perform APAPS in a compact setup.

\section{Discussion}

In this paper we have demonstrated a novel concept for performing attosecond-pump attosecond-probe spectroscopy based on a strategy including the following points: (1) The use of a commercial Ti:sapphire laser system at kHz repetition rate, (2) hollow-core fiber post-compression in a well-established parameter regime, (3a) the generation of XUV pump and probe pulses in a compact HHG setup, where the driver laser is tightly focused and where the gas jet is placed several Rayleigh lengths in front of the driving laser focus, (3b) the implementation of HHG in an overdriven regime, where the driving laser experiences an ionization-induced blueshift before the XUV generation takes place, and (4) the generation of two tightly focused attosecond pulses with different photon energies using multilayer mirrors. This concept has not only allowed us to perform an auto-correlation measurement, but also a proof-of-concept demonstration of a two-color pump-probe experiment using an attosecond pulse both in the pump and in the probe step. Our APAPS strategy has requirements which are either similar to standard attosecond experiments (regarding the footprint of the setup, the pulse compression and HHG) or which are even lower, in particular regarding the fact that a commercial, turn-key laser can be used. Therefore, our strategy can be implemented in many research laboratories around the world. As such our work places the field of attosecond science in a very similar position as the field of femtochemistry~\cite{zewail93}, which is a vibrant research field permitting detailed investigations of light-matter interactions using sequences of two or more femtosecond (rather than attosecond) laser pulses.

While the demonstrated concept allows for very stable experimental conditions, the quality of the proof-of-principle experiments that we have presented is currently limited by shot noise. There are at least three routes that we envision can be pursued in the near future to substantially improve the statistics in APAPS experiments using our scheme: (i) The HHG conversion efficiency may be further increased by tailoring the gas jet properties. It has been shown theoretically~\cite{major21b} that a homogeneous gas density is advantageous compared to the approximately parabolic gas density distribution that is present in our HHG medium~\cite{drescher18, major21}. As a result, we see potential for a substantial increase of the XUV intensity and a corresponding reduction of shot noise. (ii) The moderate pulse energy that we have used for HHG means that the scheme that we have demonstrated in this paper can be implemented using laser systems with much higher repetition rates. For instance, suitable laser systems operating at 100~kHz~\cite{kuhn17} or 1~MHz~\cite{klas21} are already available or will become available soon. These high repetition rates may even pave the route for performing attosecond-pump attosecond-probe coincidence spectroscopy~\cite{doerner00}. (iii) In the current work we have used a target density of $10^{13}$~cm$^{-3}$. While it is likely that a substantial increase of the target density will lead to problems with space charge in the case of ion detection, the target density can be increased by several orders of magnitude when performing photon-based detection as in the case of transient absorption. For instance, we anticipate that like in the afore-mentioned XUV-NIR transient absorption measurements~\cite{weisshaupt17, niedermayr22}, pump-probe signal changes on the order of $10^{-4}$ or lower may become observable in future attosecond-pump attosecond-probe transient absorption measurements. This may be aided by the relatively high photon flux that can be achieved following our approach.

Our concept differs from previous approaches using large setups and low-repetition rate laser systems~\cite{takahashi13, tzallas11, kretschmar22, driver22}, which focused on the generation of the highest possible XUV or X-ray pulse energies. In comparison, we generate a lower XUV pulse energy and intensity ($\approx 10^{13}$~W/cm$^2$) in a compact setup which allows for very stable conditions. We expect this combination to enable the application of APAPS to solid-state targets: On one hand, damage of the samples in one or a few laser shots is prevented, and on the other hand, a high sensitivity to small transient signal changes can be achieved, which is often crucial in transient absorption or reflection spectroscopy of solids.

Topics that could be addressed with our technique cover a broad spectrum, ranging from multi-electron correlation processes manifesting themselves via inner-shell decay~\cite{kissin21} and interatomic and intermolecular Coulombic decay~\cite{cederbaum97} to studies of quantum entanglement in the attosecond time domain~\cite{vrakking21, koll22}. Furthermore, APAPS could be used to measure electronic coherences and dephasing times~\cite{wituschek20, geneaux20} and thus to help to better understand charge migration in biologically relevant systems~\cite{calegari14}.

\section{Online Methods}

\subsection{Generation and characterization of few-femtosecond driving pulses}

NIR pulses were generated using a commercial laser system (Spitfire Ace, Spectra Physics) at 1~kHz with a central wavelength of 800~nm and a pulse duration of 36~fs. A fraction of 3~mJ from the total pulse energy of 13~mJ was used in the current experiments. The pulses were focused into a differentially pumped hollow-core fiber (with a length of 1~m and a core diameter of 400~$\mu$m) that was filled with He at a pressure of 3.5~bar. Temporal compression was achieved by a set of six chirped mirrors (PC70, Ultrafast Innovations) and two wedges. The pulses were characterized by spatially-encoded arrangement spectral phase interferometry for direct electric field reconstruction (SEA-F-SPIDER)~\cite{witting11}, giving a pulse duration of 3.8~fs.

\subsection{Generation of intense attosecond pulses}

After compression the NIR pulses with a pulse energy of 1~mJ were focused using a spherical mirror with a focal length of 75~cm. The focused peak intensity was $5 \times 10^{15}$~W/cm$^2$. A pulsed gas jet~\cite{irimia09} was positioned at about 3.5~cm before the driving laser focus, where the NIR peak intensity was estimated as $1 \times 10^{15}$~W/cm$^2$ (when the gas jet was switched off). The HHG yield was optimized using an iris with a diameter of 8~mm. Kr and Xe (using a backing pressure of 4~bar) were used for the two-color and the one-color APAPS experiments, respectively. We note that the parameters used in the HHG scheme were the result of an optimization process for the available laser pulses, resulting in different conditions compared to our proof-of-principle demonstration of generating intense XUV pulses in a compact setup~\cite{major21}, where laser pulses with an energy of $8-16$~mJ and a duration of 40~fs were used. The XUV spectrum was measured using an XUV spectrometer consisting of a diffraction grating and a microchannel plate / phosphor screen assembly. An Al filter with a thickness of 100~nm strongly attenuated the NIR driving pulses after HHG. A split-and-delay unit consisting of a closed-loop delay stage, a closed-loop mirror mount and two spherical mirrors was employed to vary the delay between the pump and probe pulses and to overlap the pulses in space. Spherical multilayer mirrors with a focal length of 5~cm were used which split the XUV beam in the vertical direction. Different coatings were used: Mirror A had a high reflectivity $<25$~eV, while mirror B had a reflectivity peak at 33.5~eV (see Fig.~\ref{figure_spectrum}\textbf{a}).

The XUV peak intensity was estimated according to $I_{peak}\approx 2 \times E / (\Delta t \pi \omega_{0,h} \omega_{0,v})$, where $E$ is the XUV pulse energy, $\Delta t$ is the XUV pulse duration and $\omega_{0,h}$ and $\omega_{0,v}$ are the XUV beam waist radii in horizontal and vertical directions. The XUV pulse energy at the source was estimated as 10~nJ. To this end, an XUV photodiode (AXUV100, Opto Diode) was used after attenuating the NIR pulses using one 100-nm-thick and one 200-nm-thick Al filter with a total transmission of about 10~$\%$. After transmission losses at the Al filter and reflection losses at the XUV focusing mirrors, the XUV pulse energy from mirror A was estimated as 0.12~nJ, and the XUV pulse energy from mirror B was estimated as 0.15~nJ. The XUV beam waist radii in the horizontal and vertical directions were estimated as 1.2~$\mu$m and 2.4~$\mu$m, see Supplementary Information. Furthermore, the attosecond pulse duration from the fit was used. Correspondingly, the XUV peak intensity from mirror A was estimated as $8 \times 10^{12}$~W/cm$^2$, and the XUV peak intensity from mirror B was estimated as $1 \times 10^{13}$~W/cm$^2$.

\subsection{Ion detection}

For the detection of ions in the XUV-pump XUV-probe experiments, a velocity-map imaging spectrometer~\cite{eppink97} was employed, which was operated in spatial imaging mode~\cite{stei13}. An atomic jet was generated by a second pulsed gas jet~\cite{irimia09}. The atoms reached the interaction region through a hole in the repeller electrode, and the atomic density in the experiment was estimated to be around 10$^{13}$~cm$^{-3}$. The generated ions were recorded using a microchannel plate / phosphor screen assembly in combination with a charged-coupled device (CCD) camera.

\section*{Acknowledgments}
We thank M. Krause, C. Reiter and R. Peslin for their technical support. Furthermore, we thank the Fraunhofer Institute for Applied Optics and Precision Engineering IOF in Jena for providing the simulated reflectivity curves for the XUV multilayer mirrors.

\bibliography{Bibliography}

\end{document}